\documentclass[11pt]{article}
\begin{document}

\begin{center}
{\large \bf Cooperative effects on Optical forces- Dicke's bullet}

\vskip 0.5in 
P.V.Panat and S.V.Lawande

{\it Department of Physics, University of Pune}

{\it Pune-411007, Maharashtra, India}
\end{center}

\noindent
\vspace{0.5in} 

\noindent
\centerline{\bf Abstract} 

We investigate the cooperative effects on optical forces in a system of
$N$ two level atoms confined to a volume of dimension less than
$\lambda^3$, where $\lambda$ is radiation wavelength and driven by a
coherent radiation field with a spatial profile like Laguerre-Gaussian or
ideal Bessel beam.  We show a dramatic enhancement on optical forces as
well as the angular momentum imparted to the atom by a factor of ${\rm
N^2.}$\\

 PACS number(s) 42.50 FX , 32.80 Pj

\newpage
\noindent
Recently, there is a considerable interest in cooling and trapping of
alkali atoms at low temperatures and subsequent observation of

Bose-Einstein (BEC) condensation[1,2] which has a property of a
generating laser like source of matter waves. As demonstrated by S. Inoyu
{\em et. al.,} these neutral matter waves confined in a cigar shaped BEC[3]
show a highly directional super radiant Rayleigh scattering when pumped by
an off resonant laser beam[4]. This is attributed to a cooperative effect
due to N atoms spaced to within a wavelength of the laser and is
similar (though not identical) to Dicke's superradiance[5] phenomenon. It
required nearly twenty years to observe the superradiant intensity which
is proportional to N$^2$ accompanied by a subsequent reduction in life
time[6]. There has since been a resurgence of Dicke's idea after its
observation in atomic system of two ions by DeVoe and Brewer[7]. They
observed a transition rate of two ions $(\Gamma(R))$ as a function of
separation($R$) and found that $\Gamma(R)/\Gamma_o \approx 2$ when $R
\rightarrow 0$; where $\Gamma_o$ is transition rate of a single isolated
ion[7]. Very recently, a clear experimental demonstration of optical
superradiance is demonstrated in a thin material sample by Greiner {\em
et. al.}[8].s

Dicke's superradiance is observed in many electronic systems of lower
dimension too. The systems studied are quantum dots by Brandes {\em
et. al.}[9], and theoretical possibility of the Dicke superradiance is
predicted by the same group in semiconductor heterostructure[10] where
electrons and holes in lowest Landau levels act as two level atom$-$like
system. Dicke's superradiance has been proposed as an effective means to
reduce spontaneous decay time of an excited atom to cool and manipulate
atoms with narrow width by Djotyan {\em et. al.}[11].

In many applications of condensate matter waves, one needs to handle the
motion of a single atom [12] or of an atomic built$-$up in a trap[21] or
amplification of input matter wave[13]. Individual atomic motion can also
be guided either in free space or in a metallic or dielectric
interface[14] or in the wave guide[15]. Most of these studies are confined
to manipulate single atom in external coherent field.  Of late, Gangl and
Ritsch have theoretically shown that an ensemble of two level atom,
coupled to high Q cavity mode could be cooled by considering the
interaction between the cavity field and the motion of atoms in the
ensemble[16] indicating an interest in modification of properties due to
collective effect. In the light of the resurgence of interest in studying
collective effects like superradiance, it would be interesting to
investigate the nature of optical forces in a collective system of N atoms
in Dicke state interacting with a common radiation field. The aim of this
letter is to show that cooperative effects modify optical forces
radically.

As a model of cooperative effects we consider a collection of $N$
identical two level atoms of transition frequency $\omega_o$ confined to a
volume less than $\lambda^3 $ where $\lambda $ is a wavelength of
transition and driven by a coherent c.w. laser field of frequency
$\omega_L $. Such a model was first considered by Dicke[5] to explain
superradiance. In a typical superradiance phenomenon, the intensity of
emitted radiation by the N atom system scales as ${\rm N^2}$.  This
implies that a collective decay coefficient of $N \gamma$ where $2 \gamma
$ is Einstein's $A$ coefficient.  We show in this letter that the optical
forces (dipolar and dissipative) on the above $N$ atom system also scale
as $\rm N^2 $ thereby justifying the title of this letter.  We also show
that the maximum angular momentum imparted to the system by the beam is
$N^2 l \hbar$ where $l$ is quantum number associated with the coherent
field.

The equation of motion for the reduced atomic density operator $\rho $ for
N identical two level atoms confined to a single site and driven by a
single mode c-w laser of frequency $\omega_L $ is,

\begin{equation}
i \hbar {{\partial \rho }\over {\partial t}} = [H,\rho ] 
+ i  \hbar  \gamma (2\pi_{12}\rho \pi_{21} - \rho \pi_{21}
\pi_{12} - \pi_{21} \pi_{12} \rho )
\end{equation}
where, $$\pi_{ij}^{(\alpha )} = \mid i \rangle^{(\alpha ) } 
\, ^{(\alpha)}\langle j \mid $$ 
for $\alpha ^{th} $ atom with i and j taking values 1 and 2 for level
$\mid 1>$ (ground state) and $\mid 2>$ (excited state). 2$\gamma $ is 
Einstein's A coefficient. The total $\pi $ operator is \ \ \ $\pi _{ij}=
\sum _{\alpha=1}^N \pi _{ij}^{(\alpha)}$ and has an algebra

$$[\pi _{21},\pi _{12}] = \pi _{22}- \pi _{11} \equiv 2\pi _3 $$

$$[\pi _3,\pi _{21}^n]= n\pi _{21}^n $$  

$$[\pi _3,\pi _{12}^n]=-n\pi _{12}^n $$

$$[\pi _{21},\pi _{12}^m]=-m(m-1)\pi _{12}^{m-1}+2m\pi _{12}^{m-1}\pi _3 $$

$$[\pi _{12},\pi _{21}^n]=-n(n-1)\pi _{21}^{n-1}-2n\pi _{21}^{n-1}\pi _3 $$

Hamiltonian of atoms and of field is 

\begin{equation}
H=\hbar\omega _o \pi _{22} -i\hbar(\pi _{21}\alpha f({\bf r})e^{i\theta -i\omega _Lt}-
h.c.)
\end{equation}

It is convenient to make a transformation to the frame of reference of the
applied coherent field by defining,

$${\tilde{\rho}}=\exp(i\omega _L t \pi _{22})\ \rho \exp(-i\omega _L t
\pi _{22}) .$$

The equation(1) is then changed to,

\begin{eqnarray} 
{{\partial {\tilde{\rho}}} \over {\partial t}}& = & 
- i \Delta [\pi _{22},{\tilde{\rho}}]
-[\pi _{21}
\alpha f({\bf r})e^{i\theta}-\pi _{12}\alpha^* f^*({\bf r})e^{-i\theta},
{\tilde{\rho}}] \nonumber \\
&& +\gamma(2\pi _{12} {\tilde{\rho}} \pi _{21} 
- \pi _{21}\pi _{12} {\tilde{\rho}}- {
\tilde{\rho}}\pi _{21}\pi _{12})
\end{eqnarray}

The steady state solution ${\tilde{\rho}} _{ss}$ of equation(3) is defined
by ${{\partial {\tilde{\rho}}} \over {\partial t}}=0 $ and has a form [17]

\begin{equation}
{\tilde{\rho _{ss}}}=\frac {1}{D} \sum _{m,n=0}^N a _{mn}(-g)^{-m}(-g^*)^
{-n}\pi _{12}^m\pi 
_{21}^n
\end{equation}
where,

\begin{equation}
a _{mn}={{\Gamma (m+\delta +1)\Gamma (n-\delta +1)} \over {\Gamma 
(n+1)\Gamma (m+1)
\Gamma (1+\delta )\Gamma (1-\delta )}},
\end{equation}
\begin{equation}
D=\sum _{m=0}^N a _{mm} {{{(N+m+1)!(m!)^2}} 
\over {{(N-m)!(2m+1)!}\mid 
g \mid
^{2m}}}
\end{equation}
and 

$$\Delta=\omega _o -\omega _L, 
\ \ \delta=\frac {i\Delta}{\gamma}, \ \ g=
\frac {\alpha f e^{i\theta}}{\gamma}.$$

Calculations of average values of dipolar as well as dissipative forces involve
averages  $ \langle \pi _{21}\rangle $ and $ \langle\pi _{12}\rangle $ where,
\begin{equation}
<\pi _{ij}> = Tr(\tilde{\rho}\pi _{ij} ).
\end{equation}
The trace is taken over Dicke states $$ \mid \frac {N}{2},k> $$ where
k lies between -N/2 and +N/2 and k is appropriately summed. The expressions
for dipolar force $<{\bf F_{dipole}}>$ and the dissipative force $<{\bf F_{diss}}>$ are,
\begin{equation}
<{\bf F} _{dipole}>= i\hbar {\nabla\mid\alpha f({\bf r}\mid})( e^{-i\theta}
<\pi_{21}>-e^{i\theta} <\pi_{12}> )
\end{equation}
and

\begin{equation}
<{\bf F} _{diss}>=\hbar {\mid\alpha f({\bf r})\mid}\nabla\theta(e^{-i\theta}
<\pi _{21}> + e^{i\theta}<\pi _{12}> )
\end{equation}
 
After evaluating the requisite averages in equations(8) and (9), we obtain 
expressions for forces as,
\begin{eqnarray}
\langle {\bf F}_{dipole} \rangle & = &  \frac{2 \hbar \Delta}{\gamma} 
\frac{( \nabla \mid \alpha f{(\bf r)}) \mid }{D}  \nonumber \\
&& \sum _{m=1}^N \frac{m\Gamma(m+\delta)\Gamma(m-\delta)(N+m+1)!}
{(2m+1)!(N-m)!\Gamma(1-\delta)\Gamma(1+\delta)}
 \left\{ \frac{\mid\alpha f{(\bf r)}\mid}{\gamma} 
\right\}^{-(2m-1)}
\end{eqnarray}

and
\begin{eqnarray}
<{\bf F} _{diss}> &=& \frac{2\hbar\mid\alpha f({\bf r})\mid\nabla\theta}{D} \nonumber \\
&& \sum _{m=1}^N\frac{m^2\Gamma(m+\delta)
\Gamma(m-\delta)(N+m+1)!}{\Gamma(1-\delta)\Gamma(1+\delta)(2m+1)!(N-m)!}
\left \{ \frac{\mid\alpha f({\bf r})\mid}{\gamma}\right\} ^{-(2m-1)}
\end{eqnarray}
It is easy to see that the expressions (10) and (11) reduce to the forces on
a single atom in an e.m. field when we take N=1 as[12]
\begin{equation}
<{\bf F} _{dipole}> = 2\hbar\mid\alpha f({\bf r})\mid(\nabla\mid\alpha f({\bf r})\mid)
\frac{\Delta}{\Delta^2+2\mid \alpha f({\bf r})\mid^2 + \gamma^2}
\end{equation}

and
\begin{equation}
<{\bf F} _{diss}>=2\hbar\gamma\mid\alpha f({\bf r})\mid^2\frac{\nabla\theta}{
\gamma^2+2\mid\alpha f({\bf r})\mid^2+\Delta^2}
\end{equation}

We have plotted in Fig.(1) 
$$f_{{\rm dip}}= 
{ \langle {{\bf F}}_{{\rm dipole}} \rangle 
\over 
\left[ 2 \hbar \gamma \Delta (\nabla g) / \gamma^2 \right] 
} 
$$
and 
$$
f_{{\rm diss}}=
{ \langle {{\bf F}}_{{\rm diss}} \rangle 
\over 
\left[ 2 \hbar \gamma g^2 \nabla \theta \right] 
} 
$$ 
as a function of N for  parameters $\mid \delta \mid = 1$,  
$g = 1.5$ 
and observe that $f_{diss}$ and $f_{dip}$ tend to 
a constant value for large N. 
It means that the sums in
equations (10) and (11) and the denominator D behave similarly for large N.

Also $f _{dip}$ being a difference of two terms(see eq.8) shows a peak
whose position is almost insensitive to N for given parameters but its
value decreases with increasing detuning $\delta$. Note that $f_{diss}$ is
a dimensionless quantity which tends to a value 1 ,an asymptotic limit.We have plotted similiar graphs for different values of the parameters
$ \mid \delta \mid$ and g and observe that the
smaller the value of the Rabi frequency, smaller is a value of N around
which $f _{diss}$ tends to one. It is natural to scale the quantities like
$f _{dip}$ and $f _{diss}$ by $\frac{1}{N\gamma}$ rather than $\frac {1}
{\gamma}$ since the width of super radiant pulse is N$\gamma$ and the peak
intensity behaves as $N^2\gamma$ [18]. We then see that $<{\bf F}
_{dipole}>$ and $< {\bf F} _{diss}>$ behave as $N^2$ in the reduced
variables g/N and $\frac{ \Delta}{N\gamma}$.

For large N, asymptotic analysis of equations (6),(10) and (11) can be
made following the methods developed by Lawande {\em et al} [17]. These
authors discuss asymptotic expansion of general operator averages of the
type \\  $<{\pi _{21}}^p(\pi _{22}-\pi _{11})^r{\pi _{12}}^q>$. Operator
averages involved in our case are much simpler, namely, $<\pi _{21}>$ and
$<\pi _{12}>$. The method, even though exact is complicated. The results
obtained thus can be obtained in much more simple fashion by mean field
method. We write $\pi _{12} = \pi _-$, $\pi_{21}=\pi_{+}$ and $\pi _z = (\pi _{22}-\pi
_{11})/2$ and write equation of motion for $<\pi _{\pm}>$ and $\pi _z$.
The equations involve averages like $<\pi _{\pm}\pi _z>$ which we
approximate in mean field as,

\begin{equation}
<\pi _{\pm}\pi _z>\approx<\pi _{\pm}><\pi _z>
\end{equation}
We then go over to reduced variables $ m _{\pm}= \frac{<\pi _{\pm}>
e^{\mp{ i\theta}}}{N}$, $m _o=\frac{<\pi _z>}{N}$, $\tau =N\gamma t$,
$\mu= \frac{2\alpha\mid f(\bf r)\mid}{N\gamma}$ and $\nu= \frac{2\delta}
{N\gamma}$.
The moment equations in the mean field approximation are,
\begin{eqnarray}
\frac{dm _o}{d\tau}& = & -\frac{\mu}{2}(m_+ + m_-) - 2 m_+ m_-  \nonumber \\
\frac{dm _+}{d\tau}& = & \mu m_o - \frac{i\nu m_+}{2} + 2m_ + m_o  \nonumber \\
\frac{dm _-}{d\tau}& = & \mu m _o + \frac{i\nu m _-}{2} + 2m _-m _o
\end{eqnarray}
Using steady state solutions of equations (15)  and equations (8) and (9), we get the dipole and dissipative force per particle to be
\begin{equation}
\frac{<{\bf F}{_{dipole}}>}{N}= \frac{\hbar(N\gamma)
\nu\mu(\nabla\mu)}{4\eta (1+\xi^2)}
\end{equation}
and
\begin{equation}
\frac{<{\bf F}_{diss}>}{N}= \frac{\hbar(N\gamma)\xi\mu^2\nabla\theta}{2(1+\xi^2)}
\end{equation}

where,
${\xi^2}=\frac{1}{2}[(\sqrt{\beta^2+\nu^2})+\beta]$
and
${\eta^2}=\frac{1}{2}[(\sqrt{\beta^2+\nu^2})-\beta]$
and
$4\beta =4\mu^2+\nu^2 -4$. 

We now consider an angular momentum imparted to the system of atoms by the
beams that carry angular momentum such as Laguerre-Gaussian (LG) beam or an
ideal Bessel beam.  The torque imparted will be due to the dissipative force.
With $\theta$ as a Guoy phase for the beams, we see[19],

\begin{eqnarray}
\nabla\theta & = & [\frac{krz}{z^2+z _R^2}]
{\hat{e} _r}+\frac{l}{r}\hat{e} _{\phi} \nonumber \\
&& +
\left[
\frac{kr^2}{2(z^2+z _R^2)}(1-\frac{2z^2}{z^2+z _R^2})+k+\frac{(2p+l+1)z _R}
{z^2+z _R^2}\right] \hat{e}_z
\end{eqnarray}

for LG beam and,

\begin{equation}
\nabla\theta =[\frac{l}{r}]\hat{e} _{\phi} +(k-\frac{k_{\perp}^2}{2k})
\hat{e} _z
\end{equation}

for ideal Bessel beam. The torque imparted is,

\begin{equation}
\mid<\tau>\mid=\mid<{\bf r}\times {\bf F _{diss}}>\mid
\end {equation}

The torque imparted by either beams is,

\begin{equation}
\mid<\tau>\mid= \frac{\hbar N^2\gamma\xi\mu^2l}{2(1+\xi^2)}
\end{equation}

Equations (16),(17) and (21) are principal results of this letter. It is clear 
that there is a cooperative effect on forces as well as on torque of an external beam if the number of atoms N are within some coherence
 distance like 
$\lambda $. Then such a system of atoms experiences a large force and a very
large torque and very large angular momentum-justifying the part title
Dicke's bullet.\\

Possibility of observing Dicke's bullet is best in gaseous systems.Even though, the systems of transition metal ions in solid state matrix as
used by DeVoe and Brewer [7] and by Greiner {\it et al} [8] show superradiance in most clear terms
it probably is difficult to observe Dicke's bullet in these systems .
In practice , it is impossible to arrange N atoms at a point. However, the correlations between atoms within a distance $ {c \over {  2 N \gamma}}$ can also give rise 
to superradiance. Gaseous samples of sufficiently low density ensure negligible interaction between the dipoles . Among many gaseous systems, superradiance
in Tl-Hg [20] and Cs atomic beam  was studied sometime ago. If a Laguerre -Gaussian (LG) beam of superradiant frequency falls on such a system in a direction opposite to the existing pulse immediately after extinction of the exciting pulse , then, the superradiant atom will experience a force as given by Eq. (17) and subsequent acceleration {\it a} .The c.m. velocity imparted to the radiating atoms is $ ( {{\it a} \over {N \gamma}})$ .This will Doppler shift the peak frequency 
of $ \omega$ of superradiant transition to $ ( \omega -{{ 2 \pi {\it a}}\over {\lambda N \gamma}})$ which may be observable.

In conclusion , we have shown that the dissipative force and the torque for the Dicke system scale as $ N ^{2} $ if N atoms , radiating cooperatively, are exposed
to a coherent radiation field with spatial profile such as LG beam or ideal Bessel beam.\\

{\bf Acknowledgement}  One of the authors(PVP) wishes to thank 
INSA, DST  for the financial assistance. He is also thankful to U.S. Army
for financial support under contract grant number N68171-00-M 5886.
The other author(SVL) wishes to thank CSIR for financial assistance.\\

\newpage

\centerline {\large {\bf Figure Caption}}

Figure 1 shows $ f _{dip} $ and $ f_{diss}$  as a function of number of atoms N. $f_{dip}$ and $f_{diss}$ are dimensionless as defined in the text.

{\bf References}

\begin{enumerate}

 \item M.H.Anderson, J.R.Ensher, M.R.Mathews, C.E.Weiman and E.A.Cornell,
Science {\bf 269}, 198 (1995).

 \item J.Opt. Soc. Am. B. {\bf 6} (11) (1989) Special issue edited by S.Chu and
C.Weiman.

 \item S.Inouye, A.P.Chikkaturm D.M.Stamper$-$Kurn, H.Stenger and W.Ketterle,
Science, {\bf 185}, 571 (1999).

 \item M.G.Moore and P.Meystre, Phys. Rev. Lett. {\bf 83}, 5202, (1999).

 \item R.H.Dicke, Phys. Rev. {\bf 93}, 99 (1954).

 \item N.Skribanwitz , I.Herman, J.MacGillivray and M.Feld, Phys. Rev. Lett.
{\bf 30}, 309 (1973).

 \item R.G.DeVoe and R.G.Brewer, Phys. Rev. Lett. {\bf 76}, 2049 (1996).

 \item  C.Greiner, B.Boggs and T.W.Mossberg, Phys. Rev. Lett {\bf 85}, 3793
(2000).

 \item R.Brandes, J.Inoue and A.Shimzu, Phys. Rev. Lett. {\bf 80}, 3952 (1998).

 \item Tobias Brandes, Junichi Inoue and Akira Shimizu, Physica B; Condensed
matter, {\bf 272} 341 (1999).

 \item G.P.Djotyan, J.S.Bakos, G.Demeter and Z.Soralei, Technical
Digest$-$European Quantum Electronic Conference p209, 1996, 96TH8162.

 \item L.Allen, M.Babiker, W.K.Lai and V.E.Lembessis, Phys. Rev. A {\bf 54}
4259 (1996).

 \item C.K.Law and N.P.Bigelow, Phys. Rev. A {\bf 58}, 4791 (1998).

 \item S.Al$-$Awfi and B.Babiker, Phys. Rev A {\bf 58} 2271 (1998).

 \item S.Al$-$Awfi and B.Babiker, Phys. Rev A {\bf 58} 4768 (1998).

 \item Markus Gangl and Helmut Ritsch, Phys. Rev. A {\bf 61}, 011402(R) (2000).

 \item S.V.Lawande, B.N.Jagtap and S.S.Hasan,  J. Phys. B: At. Mol. Phys.
 {\bf 14}, 4171 (1981);
 S.V.Lawande B.N.Jagtap and R.R.Puri, {\it ibid} {\bf 18},1711 (1985).

 \item Leonard Mandle and Emil Wolf,"Optical Coherence and Quantum Optics",
 p846 (Cambridge University Press 1995).

 \item S.V.Lawande and P.V.Panat, Mod. Phys. Letts. {\bf 14}, 631 (2000).

 \item Santaram Chilukuri , Appl. Phys. Lett. {\bf 34}, 284 (1979).
 
 \item H.M. Wiseman and M.J.Collet, Phys.Letts. A {\bf 202}, 246(1995).
\end{enumerate}










\end{document}